\global\def\draftcontrol{0}

   \def\versionno{ stable-wormhole}
\catcode`\@=11

\expandafter\ifx\csname draftcontrol\endcsname\relax\global\def\draftcontrol{0}
\fi

{\count255=\time\divide\count255 by 60
\xdef\hourmin{\number\count255}
\multiply\count255 by-60\advance\count255 by\time
\xdef\hourmin{\hourmin:\ifnum\count255<10 0\fi\the\count255}}
\def\draftdate{\number\month/\number\day/\number\year\ \ \ \hourmin }

\newcommand\makepapertitle{\par
  \begingroup
    \renewcommand\thefootnote{\@fnsymbol\c@footnote}%
    \def\@makefnmark{\rlap{\@textsuperscript{\normalfont\@thefnmark}}}%
    \long\def\@makefntext##1{\parindent 1em\noindent
            \hb@xt@1.8em{%
                \hss\@textsuperscript{\normalfont\@thefnmark}}##1}%
     \newpage
     \global\@topnum\z@   % Prevents figures from going at top of page.
     \@makepapertitle
     \thispagestyle{empty}\@thanks
  \endgroup
  \setcounter{footnote}{0}%
  \global\let\thanks\relax
  \global\let\makepapertitle\relax
  \global\let\@makepapertitle\relax
  \global\let\@thanks\@empty
  \global\let\@author\@empty
  \global\let\@date\@empty
  \global\let\@title\@empty
  \global\let\title\relax
  \global\let\author\relax
  \global\let\date\relax
  \global\let\and\relax
  \def\version{\let\version\@version\@gobble}
}
\def\@makepapertitle{%
  \newpage
   \ifnum\draftcontrol=1 {}
   \version\versionno
   \vskip 3em%
   \else
   \hfill\hbox to 3cm {\parbox{4cm}{\@pubnum}\hss}%
   \vskip 3em%
   \fi
   \begin{center}%
   \let \footnote \thanks
     {\LARGE {\@title}}%
     \vskip 1.5em%
     {\normalsize%\large
       \lineskip .5em%
       \begin{tabular}[t]{c}%
         \@author
       \end{tabular}\par}%
     \vskip 1.5em%
     {\@bstract}%
     \end{center}%
     \vskip 1.5em
     \@date%
   \par
}

\gdef\@pubnum{}
\def\pubnum#1{%
  \gdef\@pubnum{#1}}

\gdef\@bstract{}
\def\Abstract#1{%
  \gdef\@bstract{%
   \parbox{\textwidth-0pc}{%
   \centerline{\bf Abstract}\penalty1000%
\kern.2cm%
\noindent%\abstractfont \baselineskip=12pt
\renewcommand\baselinestretch{1.0}%
{#1}}}
}

\def\ps@paper{\let\@mkboth\@gobbletwo%
     \ifnum\draftcontrol=1
    \def\@oddfoot{\hbox to \textwidth{\tiny \versionno \hfil\tiny\draftdate}%
    \hskip -\textwidth \hbox to \textwidth{\hfil\rm\thepage\hfil}}%
     \else\def\@oddfoot{\hbox to \textwidth{\hfil\rm\thepage\hfil}}
     \fi
     \let\@evenfoot\@oddfoot
}
\def\body{\clearpage
          \pagestyle{paper}
    }
\def\@version#1{\ifnum\draftcontrol=1
\typeout{}\typeout{#1}\typeout{}
\vskip3mm\centerline{\hbox{\fbox{\normalsize{\tt DRAFT -- #1 -- }
                   {\draftdate}}}}\vskip3mm
\fi}
\let\version\@version
\long\def\eqlabel#1{\ifnum\draftcontrol=1
                    \tag@false  % there are some problems with multline without this
                    \tag*{(\theequation) \hbox to -0.2cm{\hspace{0cm}\small{#1}\hss}}
                    \refstepcounter{equation}
                    \edef\@currentlabel{\theequation}
                    \ltx@label{#1}          % use old LaTeX \label instead of new definition
                    \else
                    \label{#1}
                    \fi
                    }
\let\st@bibitem\@bibitem
\let\st@lbibitem\@lbibitem
\ifnum\draftcontrol=1
  \def\@bibitem#1{%
    \st@bibitem{#1}\a@@label{#1}\ignorespaces}
  \def\@lbibitem[#1]#2{%
    \st@lbibitem[#1]{#2}\a@@label{#2}\ignorespaces}
  \def\a@@label#1{%
    \gdef\a@lab{\smash{\normalfont\small#1}}
    \ifvmode
      \if@inlabel
        \global\setbox\@labels\hbox{%
          \llap{\a@lab\let\a@lab\relax
                \kern\@totalleftmargin\kern\marginparsep}%
          \box\@labels}%
      \fi
    \fi}
\fi
\documentclass[12pt,letterpaper]{article}
\usepackage{amsmath,amssymb,array,calc,epsfig,rotating,bm}
\usepackage[sort]{cite}
\usepackage{graphicx}
\usepackage{psfrag,verbatim}
\usepackage{xcolor}
\usepackage{hyperref}

\ifnum\draftcontrol=1
\fi
\renewcommand\baselinestretch{1.25}
\renewcommand\section{\@startsection {section}{1}{\z@}%
                                   {-3.5ex \@plus -1ex \@minus -.2ex}%
                                   {2.3ex \@plus.2ex}%
                                   {\normalfont\large\bfseries}}
\renewcommand\subsection{\@startsection{subsection}{2}{\z@}%
                                   {-3.25ex\@plus -1ex \@minus -.2ex}%
                                   {1.5ex \@plus .2ex}%
                                   {\normalfont\normalsize\bfseries}}
\renewcommand\subsubsection{\@startsection{subsubsection}{3}{\z@}%
                                   {-3.25ex\@plus -1ex \@minus -.2ex}%
                                   {1.5ex \@plus .2ex}%
                                   {\normalfont\normalsize\it}}
\renewcommand\paragraph{\@startsection{paragraph}{4}{\z@}%
                                   {-3.25ex\@plus -1ex \@minus -.2ex}%
                                   {1.5ex \@plus .2ex}%
                                   {\normalfont\normalsize\bf}}

\numberwithin{equation}{section}

\def\revise#1       {\raisebox{-0em}{\rule{3pt}{1em}}%
                     \marginpar{\raisebox{.5em}{\vrule width3pt\
                     \vrule width0pt height 0pt depth0.5em
                     \hbox to 0cm{\hspace{0cm}{%
                     \parbox[t]{4em}{\raggedright\footnotesize{#1}}}\hss}}}}

\newcommand\nxt[1]  {\\\fnxt#1}
\newcommand{\ie}{{\it i.e.,}\ }
\newcommand{\eg}{{\it e.g.,}\ }

\def\calb         {{\cal B}}

\def\calf         {{\cal F}}

\def\calm         {{\cal M}}
\def\caln         {{\cal N}}
\def\calo         {{\cal O}}
\def\calp         {{\cal P}}

\def\calv         {{\cal V}}

\def\zet          {{\mathbb Z}}

\def\del          {\partial}

\def\sqr#1#2{{\vcenter{\vbox{\hrule height.#2pt
 \hbox{\vrule width.#2pt height#1pt \kern#1pt
 \vrule width.#2pt}\hrule height.#2pt}}}}

\def\aa1{\phi}
\def\cc1{\psi}

\def\Om{\Omega}

\catcode`\@=12

\begin{document}

%%%
%%%%%% text starts here
%%%%%%%%%

\title{\bf Plumbing the wormholes of string theory flux compactifications}

\date{March 31, 2024}
%\date\today

\author{
Alex Buchel\\[0.4cm]
\it Department of Physics and Astronomy\\ 
\it University of Western Ontario\\
\it London, Ontario N6A 5B7, Canada\\
\it Perimeter Institute for Theoretical Physics\\
\it Waterloo, Ontario N2J 2W9, Canada\\
}

\Abstract{Thus far, the known wormholes in string theory connecting disjoint
boundaries represented by finite volume quotients of hyperbolic spaces
leak: they are non-perturbatively unstable towards brane-anti-brane
nucleation in the flux backgrounds that support these
wormholes. Turning on additional fluxes suppressed this instability,
but would not completely eliminate it. We present the first example of
a non-perturbatively stable wormhole in type IIB supergravity on a
warped squashed conifold with fluxes. }

\makepapertitle

\body

\version\versionno
\tableofcontents

\section{Introduction and summary}\label{intro}

Not long after the discovery of the gauge theory/string theory correspondence 
\cite{Maldacena:1997re} an intriguing phenomenon was pointed out in
\cite{Maldacena:2004rf}:
\begin{itemize}
\item Let $H_4$ be a hyperbolic space, and $\Gamma$ is a discrete subgroup
of its symmetry group $SO(1,4)$. We pick $\Gamma$ so that
a quotient $\Sigma_4\equiv H_4/\Gamma$ is a compact, smooth,
finite volume space. 
Consider $\caln=4$ $SU(N)$ supersymmetric Yang-Mills (SYM)
on $\calb=\bigcup_i \Sigma_{4,i}$
for disconnected components $\Sigma_{4,i}$. We expect that a partition
function of SYM on $\calb$ factorizes,
\begin{equation}
Z(\calb)=\prod_i Z(\Sigma_{4,i})\,.
\eqlabel{zb}
\end{equation}
\item At strong coupling, this $\caln=4$ SYM has a dual gravitational
description as type IIB supergravity on  $EAdS_5\times S^5$ with
the $EAdS_5$ boundary being $\calb$. There exist gravitational on-shell
solutions --- {\it the wormholes} --- connecting disjoint
fragments of $\calb$. The simplest example being the wormhole 
\begin{equation}
ds_{10}^2 =\biggl[
L^2\cosh^2\left(\frac rL\right)\ \left(d\Sigma_{4}\right)^2+dr^2
\biggr]+L^2\ \left(dS_5\right)^2\,,
\eqlabel{n4w}
\end{equation}
connecting the disjoint boundary components $\Sigma_4$ as $r\to \pm \infty$.
This wormhole is supported by the self-dual 5-form flux
\begin{equation}
F_5=\calf_5+\star \calf_5\,,\qquad \calf_5=16\pi (\alpha')^2\ N\ {\rm vol}(S^5)\,.
\eqlabel{5form}
\end{equation}
The radius $L$ in \eqref{n4w} is related to the 't Hooft coupling $g_{YM}^2 N$
of the SYM as $L^4=g_{YM}^2 N (\alpha')^2$.
\item The existence of the wormholes implies the failure of the
factorization property \eqref{zb}. Thus, one naturally seeks
the mechanism that invalidates such gravitational wormhole solutions.
While the specific wormhole \eqref{n4w} is perturbatively stable,
it is non-perturbatively unstable to $D3\overline{D3}$ nucleation
in the wormhole background 5-form flux \eqref{5form}.
\end{itemize}

The lightning overview of \cite{Maldacena:2004rf}
presented above stresses the fact that we are concerned
with the top-down holographic correspondence, rather than
its potential swampland \cite{Vafa:2005ui}
realization\footnote{Some other examples
of the exotic phenomena abundant in holographic toy models and absent in
string theory are the conformal ordered phases
\cite{Buchel:2020xdk,Buchel:2022zxl}.}. 
Indeed, numerous examples of wormholes,
both perturbatively stable and unstable, were
constructed in holographic toy
models\footnote{See \cite{Marolf:2021kjc}
for a recent review and references.}.
However, their physical significance is moot,
until they are embedded in string theory. 

Attempts to construct stable wormholes in string theory
in the past focused on the proposal in  \cite{Maldacena:2004rf}
to suppress the brane-anti-brane nucleation instability, by turning
on additional 3-form fluxes:
\begin{itemize}
\item  The supergravity wormhole instabilities in holographic models are associated
with the non-perturbative tachyons of the dual gauge theory. These gauge theory
tachyons come from conformally coupled scalar $\phi$ (representing nucleated $D3$ brane
motion in the radial coordinate $r\propto \phi$ in \eqref{n4w}), which are massless
in the limit of the vanishing $\Sigma_4$ curvature $-12/\ell^2$,
\begin{equation}
V_\phi=\frac{1}{12} R_{\Sigma_4}\ \phi^2\qquad \Longrightarrow\qquad m_\phi^2=\frac 16 R_{\Sigma_4}=-\frac{2}{\ell^2}\,.
\eqlabel{rev1}
\end{equation}
\item Turning on the 3-form fluxes and/or deforming the geometry of the compact manifold $\calv_5$
of the gravitational dual (the five-sphere in \eqref{n4w}) modifies the $D3$ dynamics
so that (see eq.(3.24) of \cite{Buchel:2004rr}) near the wormhole boundaries, 
\begin{equation}
m_\phi^2=-\frac{2}{\ell^2} +m_{fluxes}^2\pm m^2_{geometry}\,.
\eqlabel{rev2}
\end{equation}
Generically $m_{\phi}^2$ will vary over $\calv_5$; and while the 3-form fluxes will always provide
a positive contribution $m_{fluxes}^2$, the specific location of the probe $D3$ brane on $\calv_5$
can either increase or decrease the curvature coupling mass term \eqref{rev1}. 
\item The past work, starting with \cite{Buchel:2004rr},
was to begin with an example of $AdS_5/CFT_4$ holographic correspondence, and turn on the 3-form fluxes,
correspondingly deform the boundary $CFT_4$ by the relevant operator(s), so that $m_\phi^2>0$
over $\calv_5$ close to the wormhole boundaries. So far all such attempts failed, see \cite{Buchel:2004rr}
and Table~1 of \cite{Marolf:2021kjc}.    
\end{itemize}

In this paper we present the first example of wormhole solutions in type IIB supergravity,
connecting the disjoint $\Sigma_4$ boundary components,  that are stable with respect to
$D3\overline{D3}$ brane nucleation. The wormholes reported here were discovered accidentally,
while analyzing the cascading gauge theory on hyperbolic spaces \cite{ksworm}.
We do not have a physical understanding as to why these particular wormholes are stable, while,
 \eg the ones realizing holographic dual to $\caln=2^*$ gauge theory on $\Sigma_4$ are not \cite{Buchel:2004rr}. 

We discuss only the brane-anti-brane nucleation instability of the new wormholes:
their perturbative stability (and the stability within the consistent truncation) will be
discussed elsewhere. We note, though, that we do not expect any perturbative instabilities:
there are no distinct branches of the wormhole
solutions\footnote{Perturbative instabilities appear when for a fixed set of UV parameters ---
the relevant coupling constants of the boundary gauge theory --- there exist different
branches of the wormhole solutions \cite{Marolf:2021kjc,ksworm}. 
}; likewise, there are no disconnect solutions with the same symmetry properties
that could compete with our on-shell wormhole solutions in the gravitational path integral.

\section{Effective action for the stable wormholes}

Our starting point is the consistent truncation of
type IIB supergravity on the warped deformed conifold with fluxes 
\cite{Buchel:2010wp,Buchel:2021yay,Buchel:2022zxl}:
\begin{equation}
\begin{split}
&S_5=\frac{1}{16\pi G_5}\int_{\calm_5} {\rm vol}_{\calm_5}\ \biggl\{
R-\frac{40}{3} \left(\nabla f\right)^2-20 \left(\nabla w\right)^2-4 \left(\nabla \lambda\right)^2
-\frac 12\left(\nabla \Phi\right)^2\\
&-18 e^{-4 f -4w-\Phi}\biggl[e^{4\lambda} \left(\nabla h_1\right)^2+e^{-4\lambda} \left(\nabla h_3\right)^2\biggr]-36 e^{-4f -4w+\Phi} \left(\nabla h_2\right)^2-\calp_{flux}-\calp_{scalar}
\biggr\}\,,
\end{split}
\eqlabel{eaeh}
\end{equation}
where
\begin{equation}
\begin{split}
\calp_{flux}=&81 e^{-\frac{28}{3}f +4w-\Phi}(h_1-h_3)^2+162  e^{-\frac{28}{3}f +4w+\Phi}
\biggl[e^{-4\lambda}\left(h_2-\frac 19 P\right)^2+e^{4\lambda} h_2^2\biggr]\\
&+72 e^{-\frac{40}{3}f}\biggl[h_1(P-9 h_2)+9 h_2 h_3 +36\Om_0\biggr]^2\,,
\end{split}
\eqlabel{pflux}
\end{equation}
\begin{equation}
\calp_{scalar}=4 e^{-\frac{16}{3}f-12 w}-24e^{-\frac{16}{3}f-2w}\cosh(2\lambda)-\frac 92 e^{-\frac{16}{3}f+8w}
\biggl(1-\cosh(4\lambda)\biggr)\,.
\eqlabel{pscalar}
\end{equation}
Any solution of \eqref{eaeh} can be uplifted to a solution of type IIB supergravity
on the warped deformed conifold:
\nxt the metric is:
\begin{equation}
\begin{split}
&ds_{10}^2=\Omega^2\ ds_5^2+\Om_1^2\ g_5^2+\Om_2^2\ (g_3^2+g_4^2)+\Om_3^2\ (g_1^2+g_2^2)\,,\qquad
ds_5^2=g_{\mu\nu}(x)dx^\mu dx^\nu \,,
\\
&\Om_1=\frac13\ e^{f-4w}\,,\quad \Om_2=\frac{1}{\sqrt{6}}\ e^{f+w+\lambda}\,,\quad \Om_3=\frac{1}{\sqrt{6}}\ e^{f+w-\lambda}\,,\quad \Om^{-3}=108\ \Om_1\Om_2^2\Om_3^2\,,
\end{split}
\eqlabel{10dm}
\end{equation}
where $ds_5^2$ is the metric on $\calm_5$, and $\{g_i\}$ represent the basis of
the standard one-forms on conifold base \cite{Minasian:1999tt};
\nxt $\Phi$ is the ten-dimensional dilaton, with the asymptotic string coupling $g_s$,
$e^\Phi\longrightarrow g_s$;
\nxt the three scalars $h_i$ uplift to a R-R 3-form flux $F_3=F_3^{top}+dC_2$
and an NS-NS 3-form flux $H_3=dB_2$ as 
\begin{equation}
\begin{split}
&B_2=h_1\ g_1\wedge g_2+h_3\ g_3\wedge g_4\,,\quad C_2=h_2\ \left(g_1\wedge g_3+g_2\wedge g_4\right)\,,
\quad F_3^{top}=\frac P9\ g_5\wedge g_3\wedge g_4\,;
\end{split}
\eqlabel{3form}
\end{equation}
\nxt the self-dual R-R 5-form $F_5=\calf_5+\star \calf_5$ is
\begin{equation}
g_s\calf_5=\left(4\Om_0+h_2\ (h_3-h_1)+\frac P9\ h_1\right)\ g_1\wedge g_2\wedge g_3\wedge g_4\wedge g_5\,;
\eqlabel{5formks}
\end{equation}
\nxt the constant parameters $\Om_0$ and $P$ are related to the
numbers $N$ and $M$ of regular and fractional $D3$ branes at the tip of the
singular conifold, correspondingly,
\begin{equation}
\Om_0=\frac{\pi g_s(\alpha')^2 N}{16}\equiv \frac{L^4}{108}\,,\qquad P=\frac 92 M\alpha'\,,
\eqlabel{omp}
\end{equation}
where $L$ is the asymptotic radius of the conifold base; 
\nxt finally, $G_5$ is the five-dimensional effective gravitational
constant related to 10-dimensional gravitational constant of type IIB supergravity $16\pi G_{10}
=(2\pi)^7 g_s^2(\alpha')^4$ as follows
\begin{equation}
G_5=\frac{27}{16\pi^3}\ G_{10}\,.
\eqlabel{g5g10}
\end{equation}

We will be interested in a novel consistent truncation of \eqref{eaeh}:
\begin{equation}
P=0\,,\qquad \Phi\equiv 0\,,\qquad h_1= -h_3= h_2\equiv \frac{k}{12}\,,\qquad \lambda\equiv0 \,, 
\eqlabel{consttrunc}
\end{equation}
leading to 
\begin{equation}
\begin{split}
&S_5=\frac{1}{16\pi G_5}\int_{\calm_5} {\rm vol}_{\calm_5}\ \biggl\{
R-\frac{40}{3} \left(\nabla f\right)^2-20 \left(\nabla w\right)^2
-\frac{1}{2}e^{-4f-4w}\left(\nabla k\right)^2 -\calp\biggr\}\,,
\\
&\calp=4 e^{-\frac{16}{3}f-12 w}-24e^{-\frac{16}{3}f-2w}+\frac 92e^{-\frac{28}{3}f +4w}\ k^2
+\frac 18e^{-\frac{40}{3}f}\ (8-3k^2)^2 \,,
\end{split}
\eqlabel{wa}
\end{equation}
where without the loss of generality we set $\Om_0=\frac{1}{108}$, equivalently, $L=1$ \eqref{omp}.
The effective action \eqref{wa} describes the holographic dual to mass-deformed
Klebanov-Witten gauge theory \cite{Klebanov:1998hh}. The bulk scalars $\{k,w,f\}$
are dual,  correspondingly,  to the gauge invariant
operators\footnote{The subscript in $\calo_i$ indicates its conformal dimension,
$\dim(\calo_i)=\Delta_i$.}
$\{\calo_3,\calo_6,\calo_8\}$ of
the boundary gauge theory. 

\section{Euclidean wormhole solutions of \eqref{wa}}

We look for wormhole solutions in \eqref{wa} within the ansatz
\begin{equation}
ds_5^2=\frac{\left(\rho+\frac 1\rho\right)^2}{4 h}\ \left(d\Sigma_4\right)^2+\frac{h}{\rho^2}\
d\rho^2\,,
\eqlabel{background}
\end{equation}
where the metric warp factor $h$ and the bulk scalars  $\{k,w,f\}$ are functions of the
radial coordinate
\begin{equation}
\rho\in (0,\infty)\,.
\eqlabel{defrho}
\end{equation}
$\Sigma_4$ is a finite volume quotient of the hyperbolic space $H_4$ of curvature radius $\ell=1$.
The disconnected boundary components $\Sigma_4$  are located as $\rho \to 0$ and $\rho\to \infty$. 

From \eqref{wa} we find the following second order equations of motion:
\begin{equation}
\begin{split}
&0=h''-\frac{5(h')^2}{2h}+\frac{2 (3 \rho^2-2) h'}{\rho (\rho^2+1)}-\frac16
e^{-4 f-4 w} h (k')^2-\frac{20}{3}h (w')^2-\frac{40}{9} h (f')^2\\&-
\frac{3e^{-\frac{40}{3} f} h^2 k^4}{8\rho^2}
+\frac{(4 e^{-\frac{40}{3} f}-3 e^{-\frac{28}{3} f+4 w}) h^2 k^2}{2\rho^2}-
\frac{4(2 e^{-\frac{40}{3} f}-6 e^{-\frac{16}{3} f-2 w}+e^{-12 w-\frac{16}{3} f}) h^2}{3\rho^2}
\\&-\frac{4 (2 h^2 \rho^2+\rho^4+1) h}{(\rho^2+1)^2 \rho^2}\,,
\end{split}
\eqlabel{eq1}
\end{equation}
\begin{equation}
\begin{split}
&0=k''+k' \biggl(
-4 f'-4 w'-\frac{5h'}{2h}+\frac{5 \rho^2-3}{\rho (\rho^2+1)}\biggr)
-\frac{3 h k}{2\rho^2}\biggl(
3 e^{-\frac{28}{3} f+4 w} k^2+6 e^{-\frac{16}{3} f+8 w}\\&-8 e^{-\frac{28}{3} f+4 w}\biggr)\,,
\end{split}
\eqlabel{eq2}
\end{equation}
\begin{equation}
\begin{split}
&0=f''+f' \biggl( \frac{5 \rho^2-3}{\rho (\rho^2+1)}-\frac{5h'}{2h}\biggr)
+\frac{3}{40} e^{-4 f-4 w} (k')^2-\frac{3 (40 e^{-\frac{40}{3} f}-21 e^{-\frac{28}{3} f+4 w}) h k^2}{40\rho^2}
\\&+\frac{9e^{-\frac{40}{3} f} h k^4}{16\rho^2}
+\frac{4(5 e^{-\frac{40}{3} f}-6 e^{-\frac{16}{3} f-2 w}+e^{-12 w-\frac{16}{3} f}) h}{5\rho^2}\,,
\end{split}
\eqlabel{eq3}
\end{equation}
\begin{equation}
\begin{split}
&0=w''+\frac{1}{20} e^{-4 f-4 w} (k')^2+w' \biggl(
\frac{5 \rho^2-3}{\rho (\rho^2+1)}-\frac{5h'}{2h}
\biggr)
-\frac{3 h}{20\rho^2}\biggl(
3 e^{-\frac{28}{3} f+4 w} k^2+8 e^{-\frac{16}{3} f-2 w}\\&-8 e^{-12 w-\frac{16}{3} f}\biggr)\,,
\end{split}
\eqlabel{eq4}
\end{equation}
along with the first order constraint,
\begin{equation}
\begin{split}
&0=(k')^2+\frac{80}{3} e^{4 f+4 w} (f')^2+40 e^{4 f+4 w} (w')^2
-\frac{6 e^{4 f+4 w} (h')^2}{h^2}
+\frac{24 (\rho^2-1) e^{4 f+4 w} h'}{\rho (\rho^2+1) h}
\\&-\frac{3 (3 e^{-\frac{16}{3} f+8 w}-4 e^{-\frac{28}{3} f+4 w}) h k^2}{\rho^2}
-\frac{9e^{-\frac{28}{3} f+4 w} h k^4}{4\rho^2}
-\frac{96 e^{4 f+4 w} h^2}{(\rho^2+1)^2}
-\frac{24 e^{4 f+4 w} (\rho^2-1)^2}{(\rho^2+1)^2 \rho^2}
\\&-\frac{8 (e^{-\frac43 f-8 w}-6 e^{-\frac43 f+2 w}+2 e^{-\frac{28}{3} f+4 w}) h}{\rho^2}\,.
\end{split}
\eqlabel{eqc}
\end{equation}
We explicitly verified that \eqref{eqc} is consistent with \eqref{eq1}-\eqref{eq4}.

Note that the equations \eqref{eq1}-\eqref{eqc} respect the $\zet_2$
symmetry\footnote{The reason for the minus sign for $k$ will be clear
from the discussion of the linearized approximation.}, that exchanges the disjoint boundary
components $\Sigma_4$:
\begin{equation}
\zet_2:\qquad \rho\longleftrightarrow \frac 1\rho\qquad {\rm and}\qquad
\{k,w,f,h\}\longleftrightarrow
\{-k,w,f,h\}\,.
\eqlabel{sym}
\end{equation}
Of course, solutions do not need to respect this $\zet_2$ symmetry; however, the $\zet_2$ symmetric
wormholes are the simplest. In what follows we limit the discussion to the symmetric wormholes.

Let $m$ be the non-normalizable coefficient of the bulk scalar $k$, \ie
$k=m \rho +\cdots$ as $\rho \to 0$. Linearizing \eqref{eq1}-\eqref{eqc} to $\calo(m)$
we find,
\begin{equation}
\begin{split}
&k=m\ \frac{\rho(1-\rho^4-4\rho^2\ln\rho)}{(1+\rho^2)^3}+\calv\ \frac{\rho^3}{(1+\rho^2)^3}+\calo(m^3)\,,\\
&h=1+\calo(m^2)\,,\qquad \{w,f\}=0+\calo(m^2)\,.
\end{split}
\eqlabel{linsol}
\end{equation}
Note that the non-normalizable coefficient of $k$ at the other boundary component, \ie
as $\rho\to \infty$, is $(-m)$ --- this explains the choice of the minus sign in \eqref{sym}.
$(\calv\mp 3m)\propto \calo(m)$
are related to the one-point correlation functions of the dual operators $\calo_3$ at the
disconnected boundary components.
$\zet_2$ symmetric wormholes at the linearized order in $m$ require $\calv=0$.  

Beyond the linear order in $m$, $\zet_2$ symmetry \eqref{sym} constrains the wormhole solutions
at the neck, \ie at  $\rho=1$ (the fixed point of the radial coordinate transformation), as
\begin{equation}
\frac{d}{d\rho}\{h,w,f\}\bigg|_{\rho=1} =0\qquad {\rm and}\qquad  k\bigg|_{\rho=1}=0\,.
\eqlabel{neck}
\end{equation}
We solve \eqref{eq1}-\eqref{eqc} numerically, subject to the following asymptotics:
\nxt in the UV, \ie as $\rho\to 0$,
\begin{equation}
\begin{split}
&k=m \rho+\frac12 m h_1\ \rho^2+\biggl(k_3+\left(\frac23 m^3-4 m\right) \ln \rho\biggr)\ \rho^3
+\calo(\rho^4\ln\rho)\,,
\end{split}
\eqlabel{kbi}
\end{equation}
\begin{equation}
\begin{split}
&w=-\frac{1}{40} m^2 \rho^2+\cdots+\biggl(w_6+\calo(\ln\rho)\biggr) \rho^6+\calo(\rho^7\ln^2\rho)\,,
\end{split}
\eqlabel{wbi}
\end{equation}
\begin{equation}
\begin{split}
&f=-\frac{3}{80} m^2 \rho^2+\cdots+\biggl(f_8+\calo(\ln\rho)\biggr) \rho^8+\calo(\rho^9\ln^2\rho)\,,
\end{split}
\eqlabel{fbi}
\end{equation}
\begin{equation}
\begin{split}
&h=1+h_1 \rho+\biggl(\frac34 h_1^2+\frac16 m^2\biggr) \rho^2+\calo(\rho^3)\,,
\end{split}
\eqlabel{hbi}
\end{equation}
were we highlighted the normalizable coefficients characterizing the
solution: $h_1$ --- the rescaling  of the radial coordinate that
enforces the wormhole neck location at $\rho=1$, and $\{k_3,w_6,f_8\}$ --- determining
the expectation
values of boundary gauge theory operators $\{\calo_3,\calo_6,\calo_8\}$ correspondingly;
\nxt at the wormhole neck, \ie as $y\equiv 1-\rho\to 0$,
\begin{equation}
h=h^h_0+\calo(y^2)\,,\qquad k=k^h_1 y+\calo(y^2)\,,\qquad f=\ln f^h_0+\calo(y^2)\,,\qquad
w=\ln w^h_0+\calo(y^2)\,,
\eqlabel{ir}
\end{equation}
where we highlighted the wormhole neck parameters $\{h^h_0,k^h_1,f_0^h,w_0^h\}$.

In total, given $m$, there are $4+4$ adjustable parameters determining any $\zet_2$
symmetric wormhole, obtained as a solution of a coupled system of 4 second order ODEs
\eqref{eq1}-\eqref{eq4}. While the UV asymptotic expansions \eqref{kbi}-\eqref{hbi} 
automatically solve \eqref{eqc}, at the neck there is a nontrivial constraint:
\begin{equation}
\begin{split}
&0=-1+\frac{(k^h_1)^2}{24(f^h_0)^4 (w^h_0)^4 (h^h_0)^2}+\frac{2}{(f^h_0)^{16/3} (w^h_0)^2 h^h_0}
-\frac{2}{3 (f^h_0)^{40/3} h^h_0}-\frac{1}{3 (f^h_0)^{16/3} (w^h_0)^{12} h^h_0}\,.
\end{split}
\eqlabel{const}
\end{equation}
In all the numerical
solutions\footnote{As an independent check, these solutions
were also constructed in a different radial coordinate gauge,
more suitable for the analysis of the cascading gauge theory on
$\Sigma_4$ \cite{ksworm}. There is a perfect agreement.}
reported, the constraint \eqref{const} is satisfied at
$\sim 10^{-6}$ (in fact, much better for smaller values of $m$),
relative to its first terms $(-1)$.

\begin{figure}[ht]
\begin{center}
\psfrag{m}[c]{{$m$}}
\psfrag{d}[cb]{{$d^2$}}
\psfrag{k}[ct]{{$\frac{k_3}{m}+3$}}
  \includegraphics[width=3.0in]{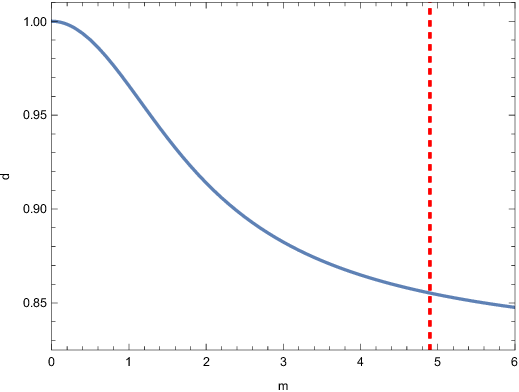}
  \includegraphics[width=3.0in]{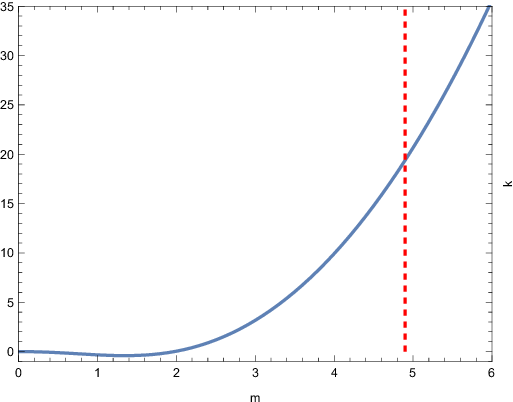}
\end{center}
\caption{Left panel: the size of the wormhole neck $d^2$ (see \eqref{defd})
as a function of the mass parameter $m$. Right panel: the normalizable
coefficient $k_3$ of the wormhole as a function of $m$. The red dashed vertical
lines indicate $m_{crit}^2=24$. Wormholes with $m^2> m_{crit}^2$ are
stable with respect to $D3\overline{D3}$ nucleation (see section \ref{sw}).
}\label{figure1}
\end{figure}

In fig.~\ref{figure1} we present the results of the numerical analysis.
In the left panel we plot the ``size of the wormhole neck'' $d$, defined
as the warp-factor of $\Sigma_4$ in 10d metric \eqref{10dm}, evaluated at $\rho=1$,
as a function of $m$ :
\begin{equation}
d^2\equiv\ \Om^2\cdot \frac{\left(\rho+\frac 1\rho\right)^2}{4 h}\bigg|_{\rho=1}=\frac{e^{-\frac{10}{3}f}}{h}
\bigg|_{\rho=1}\,.
\eqlabel{defd}
\end{equation}
In the right panel we plot the values of $(k_3/m+3)$ as a function of $m$.
The dashed vertical lines correspond to $m^2_{crit}= 24$.
As we show in section \ref{sw}, wormholes with $m>m_{crit}$ are stable
with respect to $D3\overline{D3}$ nucleation.

\section{Non-perturbatively stable wormholes}\label{sw}

Wormholes are stable with respect to the nucleation of  $D3\overline{D3}$
pairs if the potential for the probe $D3$ is confining --- there is
no run-away behavior --- near the wormhole boundaries.
Indeed, otherwise, the $D3$ branes nucleated by the background 5-form
flux will move to the wormhole boundaries, reducing this flux
and ultimately leading to the collapse of the wormhole. 
As we remind the reader below, the nucleated $\overline{D3}$ branes
will move to the wormhole neck. 

The effective action for the probe $D3/\overline{D3}$ brane in generic type IIB
supergravity background was studied in \cite{Buchel:2004rr}.
For a 3-brane of charge $q$ (we use $q=+1$ for a $D3$ brane and
$q=-1$ for an $\overline{D3}$ brane) with a world-volume along $\Sigma_4$
the only nontrivial dynamics occurs in the radial direction (the motion
along the base of the squashed conifold is flat):
\begin{equation}
S_{3-brane}=-T_3 \int_{\Sigma_4} d^4 x \sqrt{\hat g}+q T_3 \int_{\Sigma_4} C_4\,,
\eqlabel{s3}
\end{equation}
where $T_3$ is the 3-brane tension; $\hat g$ is the induced
metric on the world-volume of the probe, see \eqref{10dm} 
with $ds_5^2$ given by \eqref{background},
\begin{equation}
\hat g_{\mu\nu}=\frac{\Omega^2(\rho+\frac 1\rho)^2}{4h}\ \tilde{g}_{\mu\nu}+
\frac{\Omega^2 h}{\rho^2} \del_\mu\rho \del_\nu\rho\,,
\eqlabel{defgh}
\end{equation}
where $\tilde{g}_{\mu\nu}$ is the metric on $\Sigma_4$; 
and, see \eqref{5formks} with \eqref{consttrunc},
\begin{equation}
\begin{split}
dC_4=\star \calf_5 =& -\left(4\Omega_0-\frac{k^2}{72}\right)\cdot \Omega_1^{-1}\Omega_2^{-2}
\Omega_3^{-2}\cdot \Omega^5\frac{(\rho+\frac 1\rho)^4}{(4 h)^2}\ \frac{h^{1/2}}{\rho}\
{\rm vol}(\Sigma_4)\wedge d\rho\\
=&{\rm vol}(\Sigma_4)\wedge d\omega_4\,,
\end{split}
\eqlabel{c4}
\end{equation}
where the last line defines the zero form $\omega_4(\rho)$. 
Recalling $\Omega_0=\frac{1}{108}$ and using \eqref{10dm}, we find for slowly varying
$\rho(x^\mu)$,
\begin{equation}
S_{3-brane}=T_3\int_{\Sigma_4}\sqrt{\tilde{g}}\biggl(-\frac 12\cdot
\frac{\Omega^4(1+\rho^2)^2}{4\rho^4}(\del\rho)^2-V_q(\rho) \biggr)\,,
\eqlabel{effac}
\end{equation}
with the effective potential
\begin{equation}
V_q(\rho)=\frac{\Omega^4(\rho+\frac{1}{\rho})^4}{16 h^2}-q \omega_4\,.
\eqlabel{defepot}
\end{equation}
Since we are interested in the effective 3-brane action
near the wormhole boundaries, we can use the asymptotic expansions \eqref{kbi}-\eqref{hbi}.
We find:
\begin{equation}
\begin{split}
\omega_4=\frac{1}{16\rho^4}-\frac{h_1}{8\rho^3}+\frac{1}{\rho^2}
\biggl( \frac{3}{32} h_1^2-\frac{1}{64} m^2+\frac12\biggr)
+\frac{1}{\rho}\biggl(
-\frac{1}{32} h_1^3+\frac{1}{64} m^2 h_1-\frac38 h_1\biggr)+\calo(\rho)\,,
\end{split}
\eqlabel{defw4}
\end{equation}
leading to
\begin{equation}
V_q=\frac{1-q}{16\rho^4}+\frac{(q-1)h_1}{8\rho^3}
+\frac{1}{\rho^2}\biggl(\frac{3}{32}(1-q) h_1^2+\biggl(-\frac{1}{192}+\frac{q}{64}\biggr) m^2
+\frac14-\frac q2
\biggr)+\calo(\rho^{-1})\,.
\eqlabel{vfin}
\end{equation}
It is convenient to introduce a canonically normalized scalar $\phi$ according to,
see \eqref{effac},
\begin{equation}
\begin{split}
d\phi \equiv& -\frac{\Omega^2(1+\rho^2)}{2\rho^2}\ d\rho\\
=&\frac{1}{2\rho}+\left(-\frac12-\frac{1}{16}m^2\right)\ \rho-\frac{1}{32}h_1 m^2\ \rho^2+\calo(\rho^3\ln\rho)\,,
\end{split}
\eqlabel{phir}
\end{equation}
resulting in
\begin{equation}
S_{3-brane}=T_3\int_{\Sigma_4}\sqrt{\tilde{g}}\biggl(-\frac 12 (\del\phi)^2-V_q(\phi)\biggr)\,,
\eqlabel{s3fin}
\end{equation}
with the scalar potential 
\begin{equation}
V_q(\phi)=(1-q)\ \phi^4+h_1 (q-1)\ \phi^3+\biggl(
\left(\frac{5}{48}-\frac{q}{16}\right) m^2+2-3 q-\frac38 h_1^2 (q-1)\biggr)\ \phi^2+\calo(\phi)\,,
\eqlabel{vfinphi}
\end{equation}
valid for large values of $\phi$. 
From \eqref{vfinphi}, the $\overline{D3}$ potential is always confining (causing the nucleated
anti-brane to move to the wormhole neck)
$V_{-1}\propto +\phi^4$, while the $D3$ potential, 
\begin{equation}
V_{+1}=\biggl(\frac{m^2}{24}-1\biggr)\ \phi^2+ \calo(\phi^0)\,,
\eqlabel{q1}
\end{equation}
is confining only provided
\begin{equation}
m^2\ell^2\ >\  m_{crit}^2\ell^2 \equiv \frac{1}{24}
\eqlabel{mcritdef}
\end{equation}
where we restored the proper dimensions using the $\Sigma_4$ length scale $\ell$. 

Eq.~\eqref{mcritdef} is the stability criteria with respect to $D3\overline{D3}$
pair nucleation in $\zet_2$-symmetric wormholes of the effective
action \eqref{wa}.

\section*{Acknowledgments}
Research at Perimeter
Institute is supported by the Government of Canada through Industry
Canada and by the Province of Ontario through the Ministry of
Research \& Innovation. This work was further supported by
NSERC through the Discovery Grants program.

\bibliographystyle{JHEP}
\bibliography{sworm}

\end{document}